\documentclass[journal=jcisd8,manuscript=article]{achemso}
\usepackage[version=3]{mhchem} 
\usepackage{multirow}
\usepackage{multicol}
\usepackage{setspace}
\usepackage{hyperref}

\author{Alejandro Varela-Rial}
\affiliation{Acellera Labs, Doctor Trueta 183, Barcelona, Spain}
\alsoaffiliation{Computational Science Laboratory, Universitat Pompeu Fabra, Barcelona Biomedical Research Park (PRBB), Barcelona, Spain}
\author{Maciej Majewski}
\affiliation{Computational Science Laboratory, Universitat Pompeu Fabra, Barcelona Biomedical Research Park (PRBB), Barcelona, Spain}
\author{Alberto Cuzzolin}
\affiliation{Acellera Labs, Doctor Trueta 183, Barcelona, Spain}
\author{Gerard Mart\'inez-Rosell}
\affiliation{Acellera Labs, Doctor Trueta 183, Barcelona, Spain}
\author{Gianni De Fabritiis}
\email{gianni.defabritiis@upf.edu}
\affiliation{Computational Science Laboratory, Universitat Pompeu Fabra, Barcelona Biomedical Research Park (PRBB), Barcelona, Spain}
\alsoaffiliation{Acellera Labs, Doctor Trueta 183, Barcelona, Spain}
\alsoaffiliation{Instituci\'o Catalana de Recerca i Estudis Avan\c{c}ats (ICREA), Passeig Lluis Companys 23, Barcelona, Spain}

\title[SkeleDock]
  {SkeleDock: A Web Application for Scaffold Docking in PlayMolecule}

\keywords{Docking, D3R, MCS, \LaTeX}

\begin{document}
\setstretch{1.25}
\begin{abstract}

SkeleDock is a scaffold docking algorithm which uses the structure of a 
protein-ligand complex as a template to model
the binding mode of a chemically similar system. This algorithm
was evaluated in the D3R Grand Challenge 4 pose prediction challenge, where it achieved competitive performance. Furthermore, we show that, if crystallized fragments of the target ligand are
available, SkeleDock can outperform rDock docking software at 
predicting the binding mode. This article also addresses
the capacity of this algorithm to model macrocycles and deal
with scaffold hopping. SkeleDock can be accessed at https://playmolecule.org/SkeleDock/. \\

\end{abstract}

\begin{small}
\begin{multicols}{2}
\section{Introduction}
Predicting the binding mode of small molecules in a protein pocket is one of the main challenges in the field of computational chemistry. Accurate predictions can substantially reduce the costs of drug development and speed up the process\cite{Pinzi2019}. Several software solutions exist that address this problem, including AutoDock Vina \cite{Trott2010}, Glide \cite{Friesner2004}, Gold \cite{Jones1997} or rDock \cite{Ruiz-Carmona2014}. Typical docking protocols use the protein cavity and the query ligand to generate poses that are later evaluated with a generalized scoring function. However, structural knowledge about the target system is usually available, such as protein homologs with similar co-crystallized ligands. Hence, given that the binding mode of similar molecules is usually conserved \cite{Drwal2017, Malhotra2017}, it is reasonable to exploit this information to increase the accuracy of the prediction.  Considering the growing amount of structural data available in the Protein Data Bank (PDB)\cite{Berman2000}, and the popularity of fragment-based drug discovery \cite{Lamoree2017}, we expect this knowledge-rich scenario to become increasingly prevalent. \\

Docking algorithms which make use of such knowledge are usually referred to as similarity docking or scaffold docking \cite{Fradera2004}. Scaffold docking methods usually rely on maximum common substructure (MCS) approaches, such as fkcombu \cite{Kawabata2014}. MCS methods try to find the largest common substructure (subgraph) between two molecules. When found, the conformation of that substructure in the query ligand can be modelled by simply mimicking the conformation of that same substructure in the template, while the position of the remaining atoms is decided by a general scoring function. However, due to the characteristics of the MCS methods, two almost identical molecules that only differ in minor modifications, can return disappointingly short subgraphs. A shorter MCS means that more atoms in the query ligand would have to be modelled without any reference by the docking software, which is not desirable. Additionally, such minimal mismatches can be of critical interest in medicinal chemistry, as they can constitute scaffold hops that can, potentially, improve the pharmacological properties of a compound or circumvent intellectual property \cite{Hu2017}. Therefore, there is a need to maximize the use of structural information. We present here SkeleDock, a new scaffold docking algorithm that can overcome local mismatches. \\


\section{Features}

\begin{figure}[H]
    \includegraphics[scale=0.60]{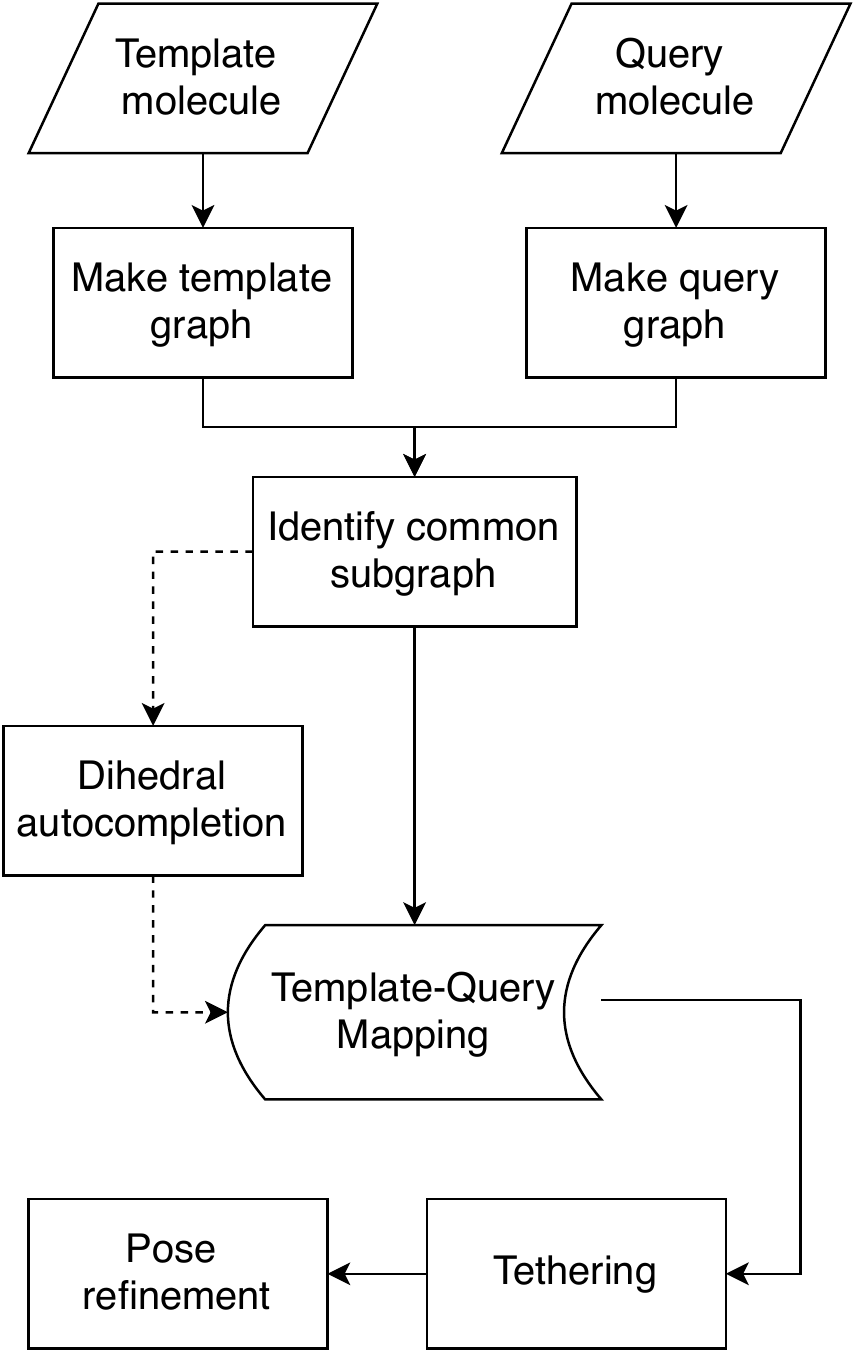}
    \caption{Main steps of SkeleDock algorithm. The dihedral autocompletion step is optional.}
    \label{skeledock_flow}
\end{figure}

\textbf{Algorithm.}
SkeleDock web application provides a user-friendly interface to perform scaffold docking, starting from files with the structure of the receptor (PDB), a template molecule (PDB) and a set SMILES representing query ligands (CSV). After submission, these files follow SkeleDock's algorithm, whose main steps are summarized in Figure \ref{skeledock_flow}. The algorithm begins by building a graph for the query and the template molecules. These two graphs are then compared to identify a common subgraph, that is, a continuous set of atoms whose element (node) and bonds (edges) are equivalent in both the query and the template molecules. Hence, if this step is successful, a mapping linking several atoms in the query molecule to their template counterparts will be returned. In the following step, tethering, this mapping is used to change the position of the atoms of the query molecule. This is done by creating a force in each query atom that points towards the location of its template counterpart, effectively biasing the conformation of the query ligand towards that of the template. Finally, in order to find an appropriate location for those atoms in the query molecule for which no template equivalent was found, the \textit{tethered template docking} protocol of rDock \cite{Ruiz-Carmona2014} is used. This protocol allows the user to constrain the degrees of freedom of the docking run (orientation, position and dihedral angles of the ligand), based on the initial conformation of the provided molecule and a set of atom indexes. These indexes correspond to the atoms that the user wants to be fixed, in our case, those atoms for which we have found a template counterpart. If a given dihedral is composed by atoms whose indexes belong to this set, its dihedral angle will not be sampled at all or only within a user-defined range. \\

\textbf{Autocompletion step.}
As previously discussed, one limitation of methods based on MCS is its sensitivity to small changes: two molecules which are almost identical, except for some minor modifications, will return a smaller mapping, as the common subgraph shared by both is now smaller. Figure \ref{dihedral_autocompletion}a depicts such scenario. To avoid this problem, we added an optional step called dihedral autocompletion. As shown in Figure \ref{dihedral_autocompletion}b, the mapping found in the graph comparison step has \textit{stopped} just before the atom whose element differs between the query and template molecules, depicted as X in Figure \ref{dihedral_autocompletion}a. However, this mismatching atom belongs to a dihedral (highlighted in a ball-stick representation) in which three consecutive atoms are already mapped to the template. We can then assume that the mismatching atom -the fourth atom of this dihedral- matches the fourth atom of the equivalent dihedral in the template. This is what we refer to as dihedral autocompletion. After each dihedral autocompletion cycle, a new non-mapped fourth atom appears, and this step is repeated recursively until no more atoms are available. If the template dihedral offers several possibilities for the fourth atom, all of them are explored and evaluated. This functionality is key to overcome local mismatches, which makes SkeleDock able to handle some minor scaffold hopps. Some MCS methods can overcome simple mismatches as the one shown in Figure \ref{dihedral_autocompletion}, as they could identify the two disconnected, common subgraphs. However, if these disconnected subgraphs are not highly similar, typical MCS methods could fail.\\

\begin{figure}[H]
    \includegraphics[scale=0.12]{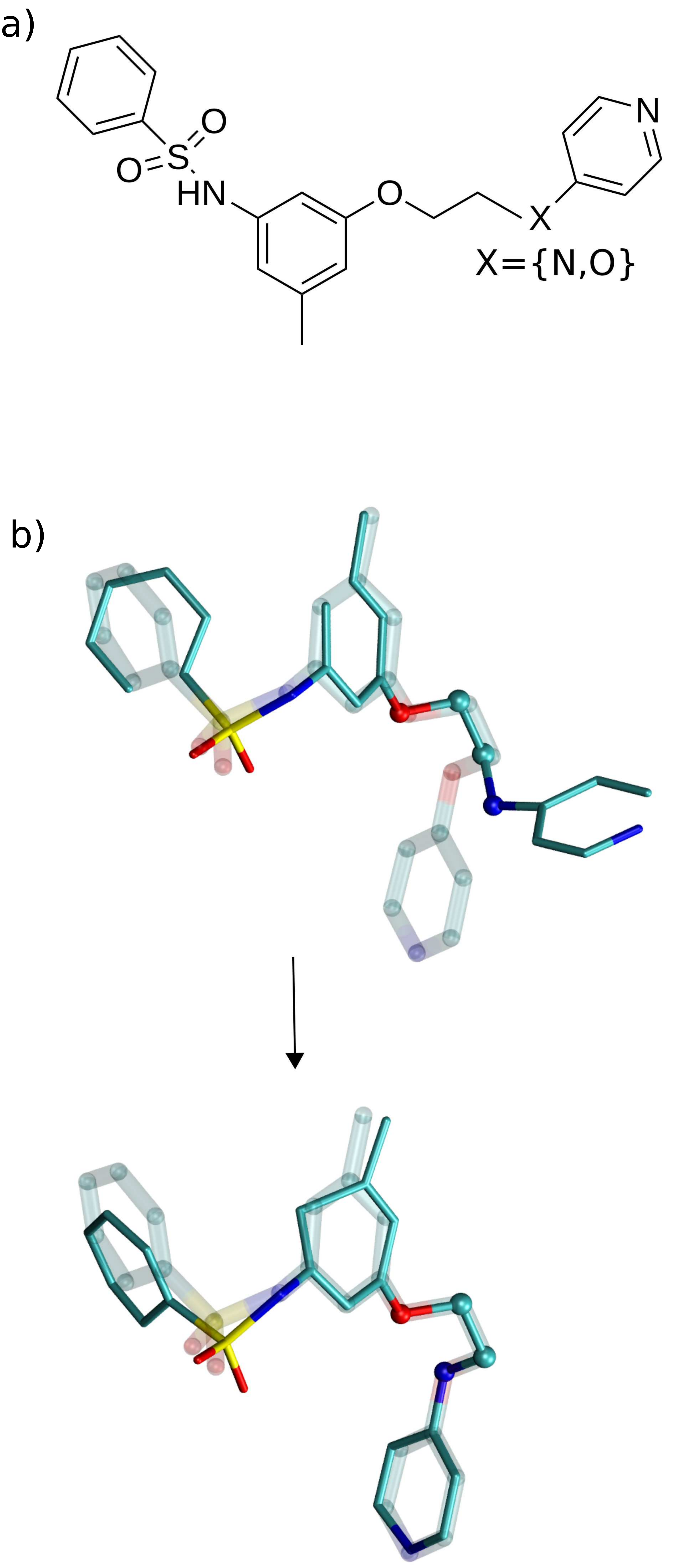}
    \caption{Dihedral autocompletion step. a) Chemical structure of template and query molecules. The mismatching atom is depicted as X. b) Overlap between query molecule (opaque licorice) and template molecule (transparent texture) before (top image) and after (bottom image) the autocompletion step. The semi-completed dihedral (atoms depicted with a ball) propagates to the right side, improving the overlap with the template. Template molecule is PDB code: 1UVT, resname: I48. Rings have to be broken to allow this step, but they are restored before the tethering. These conformations are not the final docked poses.}
    \label{dihedral_autocompletion}
\end{figure}

\textbf{Application options.}
Different options are available to change the behaviour of the application. \textit{rDock refinement step} is enabled by default, but it might not be necessary if every query atom has a template equivalent. The \textit{scaffold-hopping tolerant mode} enables or disables the dihedral autocompletion step. Users can also modify the magnitude of the force applied to each atom during tethering. Higher values result in a better alignment but might introduce some artefacts, like a change of chirality. The last option is \textit{probe radius} that defines the radius of the spheres used to define the size of the docking cavity for rDock.\cite{Ruiz-Carmona2014} After execution, the best pose of each ligand can be displayed together with the protein and the template ligand (Figure \ref{PM_GUI}). Results can be downloaded in a tar.gz file. \\

\end{multicols}
\end{small}

\begin{figure}[H]
    \includegraphics[scale=0.24]{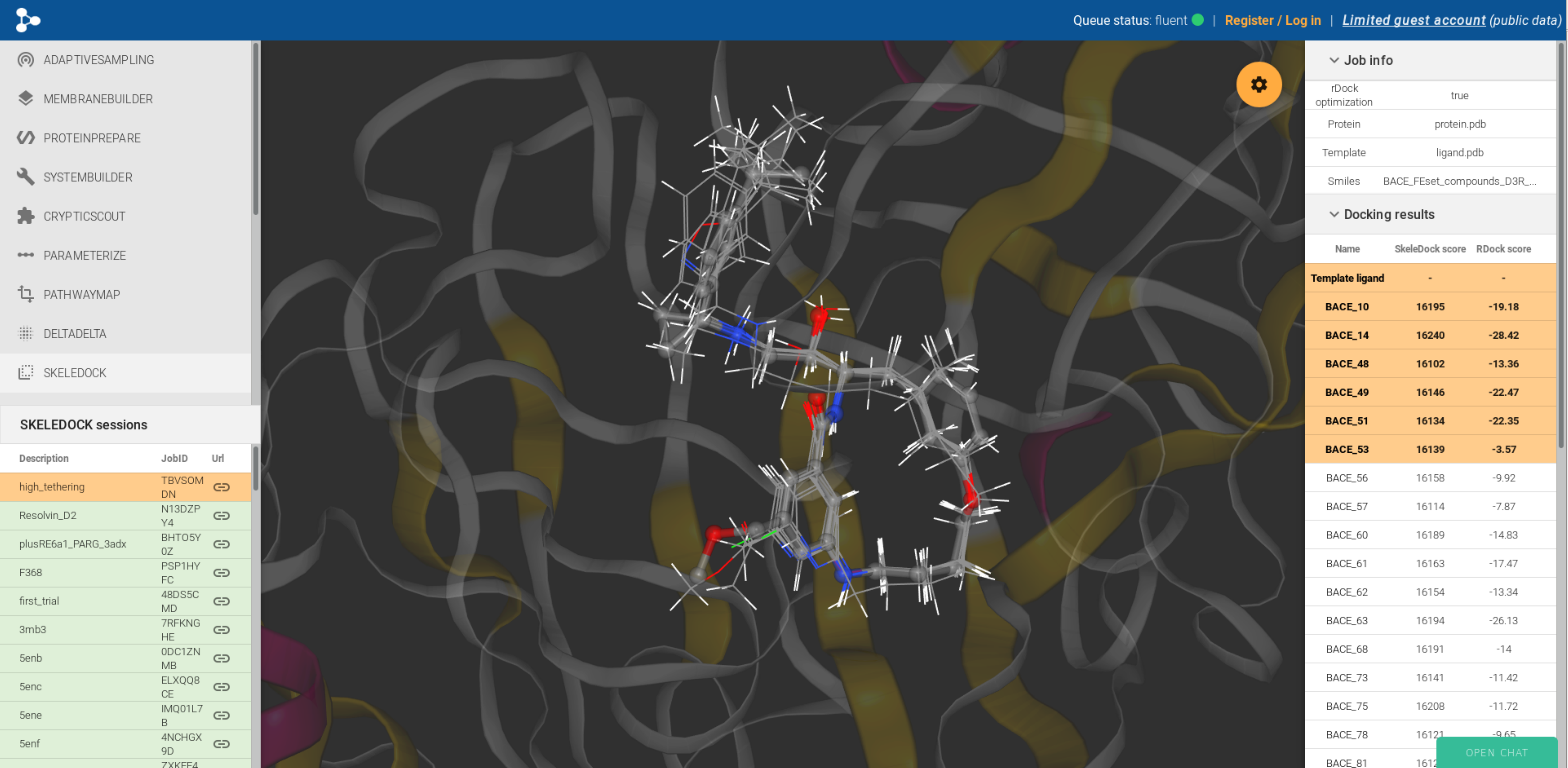}
    \caption{SkeleDock's graphical user interface. Docked molecules (line representation) are shown overlapped with the template used (ball and stick representation).}
    \label{PM_GUI}
\end{figure}

\begin{small}
\begin{multicols}{2}

\textbf{Time performance}.
We assessed the efficiency of SkeleDock by docking congeneric series for two different targets: Cathepsin S (459 ligands, average of 46.6 heavy atoms) and  BACE-1 (154 ligands, an average of 38.4 heavy atoms). We used rDock as a baseline, and each test was run using 4 and 30 cores. SkeleDock is two to three times slower than rDock, but we believe that the increase in accuracy compensates it. Table \ref{time-performace} sums up the results of time performance evaluation.\\

\begin{table}[H]
\small
 \begin{tabular}{|c c | c | c |} 
 \hline
 
\multicolumn{2}{c}{  } & \multicolumn{1}{c}{BACE-1} & \multicolumn{1}{c}{CatS}\\
\multicolumn{2}{c}{  } & \multicolumn{1}{c}{(154)} & \multicolumn{1}{c}{(459)}\\

 \hline
 
\textbf{Method} & \textbf{\#cpu} & \textbf{Yield} & \textbf{Yield} \\
 &  & [Lig/min] & [Lig/min] \\

 \hline

\multirow{2}{*}{SkeleDock} & 4  &  15.2 & 13.9 \\
                           & 30  & 43.8  & 50.7 \\

 \hline
 
 \multirow{2}{*}{rDock}    & 4   & 50.5  &  53.4  \\
                           & 30  & 102.0 & 127.5 \\

 \hline
 
 \end{tabular}
 \caption{Time performance of SkeleDock and rDock for BACE-1 and CathepsinS congeneric series. The number of simulated ligands is listed in brackets.}
\label{time-performace}
\end{table}


\section{Validation}

\textbf{Fragment-based docking.}
We evaluated SkeleDock's ability to recover the native pose of a ligand using a fragment as a template. Due to the lack of crystal structures of protein with ligands and corresponding fragments, we decided to artificially generate fragments for complexes from the refined set of PDBbind (version 2018) \cite{Wang2005} and use them as templates for SkeleDock. The ligands were fragmented by breaking a selected rotatable bond. We prepared three sets of fragments of increasing difficulty by excluding from the fragment 1, 3 or 5 rotatable bonds of the complete ligand. Deleting more atoms from the template increases the difficulty of predicting the right pose, as there is no reference for them. \\

We compared SkeleDock's performance with two MCS-based methods and with an unconstrained docking protocol. The MCS-based methods are two different settings of RDKit's \cite{rdkit} findMCS function: The first, where the element of the atoms must match (strict MCS) and the second, where the element and bond-order mismatches are allowed (agnostic MCS). The function returns a mapping (just as the graph comparison step of SkeleDock), which is then directly passed as an input to the tethering and pose refinement steps. Finally, for the unconstrained docking protocol, we used rDock with free rotation, translation and dihedral angle exploration (free rDock). The performance of docking algorithms is evaluated by the number of correct predictions. By convention, poses are considered correct if their RMSD from their crystal pose is under 2.0{\AA} \cite{Cross2009}. We report two levels of success: Top 1, where only the top pose was selected, and Top 5, where the best among the top five poses was selected (Figure \ref{fragment-based}). A full report of the docking results can be found in Table S1. \\

In terms of success rate, SkeleDock outperforms other approaches in all fragmentation scenarios (Figure \ref{fragment-based}).  Strict MCS is comparable to SkeleDock, and they both outperform agnostic MCS and free rDock. This result is expected because the less-strict nature of the agnostic MCS setting might find mappings which are feasible in terms of equivalence in other features (as ring-ring), but lead to wrong orientations of the ligand. These results suggest that, when the binding mode of the query ligand and its fragments are conserved, biasing the prediction using SkeleDock or MCS approaches can substantially increase the success rate of binding mode prediction.\\

\begin{figure}[H]
    \includegraphics[scale=0.55]{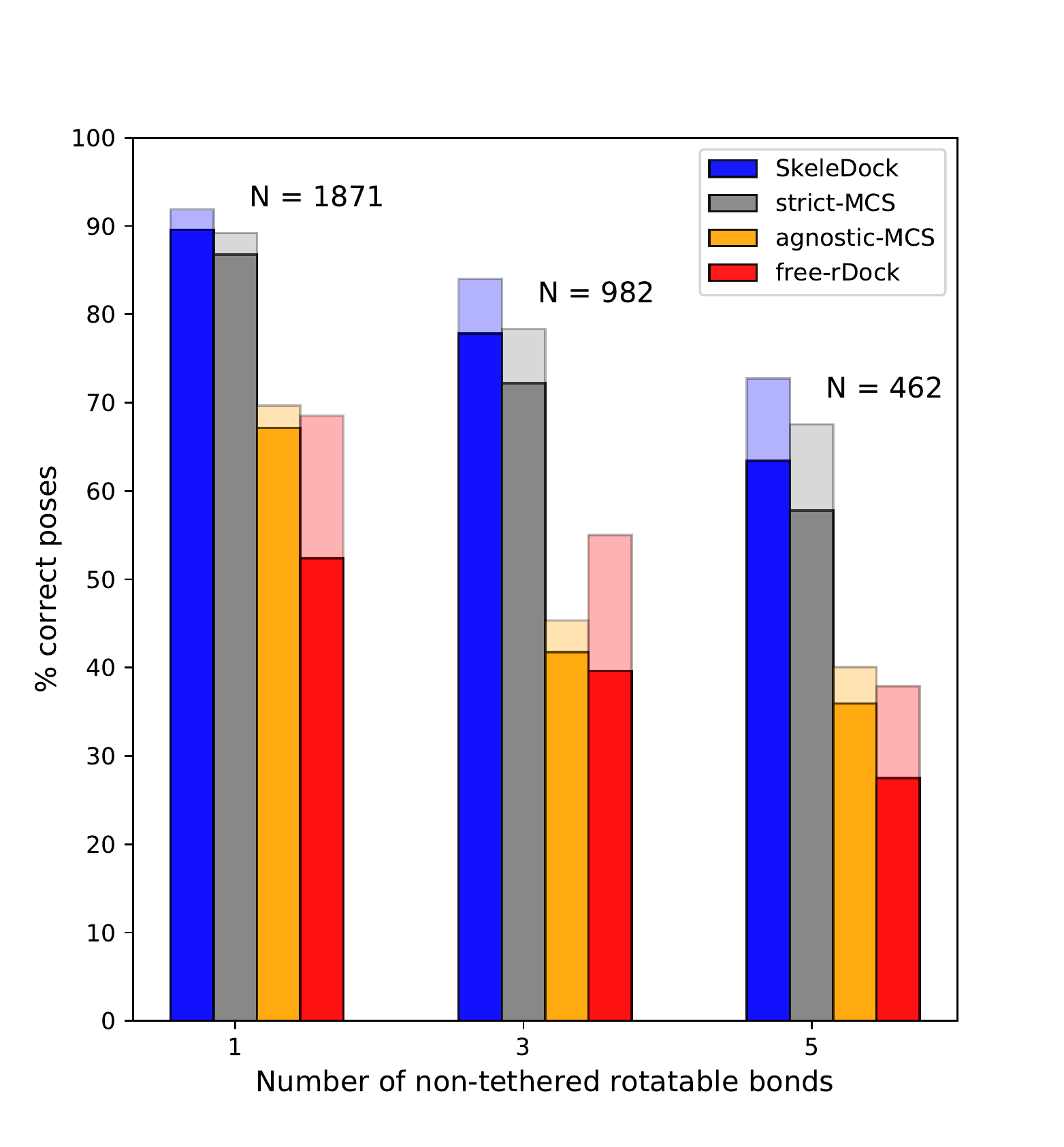}
    \caption{Self-docking performance of SkeleDock, strict MCS, agnostic MCS and free rDock. Different shades correspond to the different success levels: Top 1 - opaque, Top 5 - transparent.}
    \label{fragment-based}
\end{figure}

\textbf{D3R Grand Challenge 4.} 
In order to evaluate SkeleDock prospectively, we engaged in the D3R Grand Challenge 4 (GC4) pose prediction subchallenge. The D3R Grand Challenge is an international contest where participants complete different computational tasks of pharmaceutical interest.\cite{Gaieb2019} In its fourth edition, the objective was to predict binding modes of 20 ligands of BACE-1 protein. As templates for SkeleDock, we used crystal structures of close homologs and their co-crystallized ligands from PDB (Table S2). At the time the challenge took place, the final rDock pose refinement step was not implemented in the protocol. Instead, we run a short molecular dynamics (MD) simulation to relax the poses (See SI:MD simulation for further details). To asses the performance of the final protocol, a retrospective analysis was run using SkeleDock's web application at PlayMolecule. \\

This subchallenge was particularly complicated for two reasons: (1) all ligands except one had a macrocycle, and (2) most of the ligands had a shortened MCS with their template due to certain atoms differing in element, or the presence of rings. Conformational changes in macrocycles involve the concerted rotation of several dihedrals, making them difficult to model \cite{Allen2016}. The gold standard among docking practitioners is to first sample different conformations of a macrocycle and then dock each one independently. This was not needed in our case, as SkeleDock can simply use the macrocycle of the template to model the one in the query ligand. Regarding the shortened MCS, the autocompletion step of SkeleDock can handle these mismatches, leading to a greater mapping and overlap with the templates both in the macro and non-macro fractions of the molecules, as can be seen in Figure \ref{MCS_comparison}. We actually compared the RMSD of the poses generated by SkeleDock and the two MCS methods described in \textit{fragment-based docking}. Both the global RMSD and the macrocycle RMSD is lower in SkeleDock poses, thanks to the bigger mapping with the template (Table S3 and Table S4).\\

SkeleDock's submission (code: qqou3) finished among the top-performing participants, ranking 9th out of 74 according to median RMSD (1.02 \AA) and 15th according to mean RMSD (1.33 \AA). In the retrospective analysis, SkeleDock web application performed slightly worse with a mean RMSD of 1.47 {\AA}. Given that this test was run in a fully automated fashion and with no human supervision, the gap between the two results is understandable.\\

\begin{figure}[H]
    \includegraphics[scale=0.11]{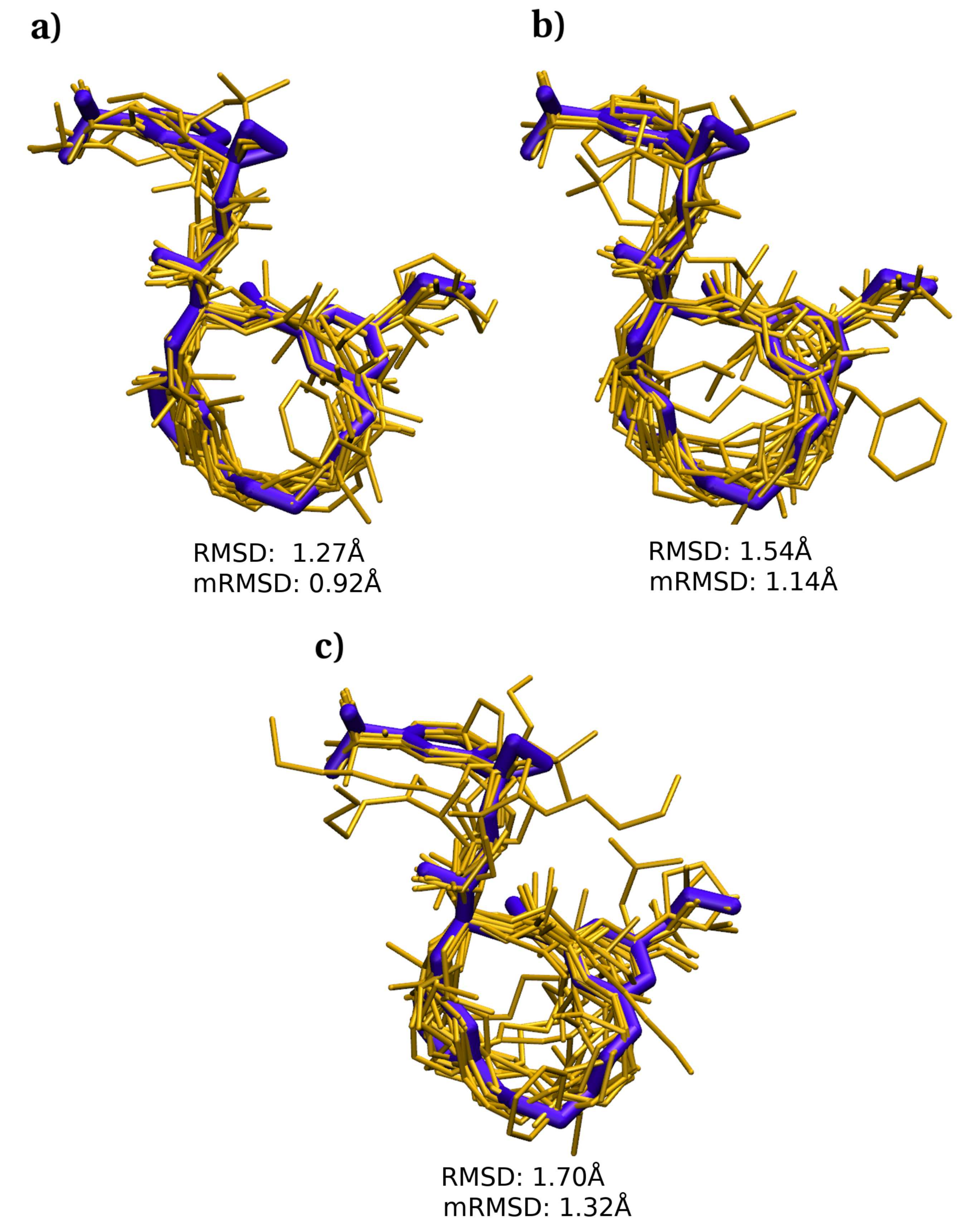}
    \caption{Overlap between predicted poses (gold) and template (violet) using 3 different methods: a) SkeleDock, b) Element Agnostic, c) Strict MCS. RMSD is the average RMSD value of the poses, and mRMSD is the mean value of the RMSD of the macrocycle atoms.}
    \label{MCS_comparison}
\end{figure}

\section{Conclusions}

SkeleDock algorithm offers four main features: 1) docking of molecules based on their analogues or fragments, 2) autocompletion step that can handle local mismatches and hence, model minor scaffold hopps, 3) ability to model macrocycles without having to pregenerate ring conformations and 4) a user-friendly GUI that enables efficient scaffold docking and results exploration. The protocol can be accessed at https://playmolecule.org/SkeleDock/. \\



\begin{acknowledgement}

The authors thank Acellera Ltd. for funding and the D3R organizers for their efforts.  G.D.F. acknowledges support from MINECO (BIO2017-82628-P) and FEDER. This project has received funding from the European Union's Horizon 2020 research and innovation programme under grant agreement No 823712 (CompBioMed2) and from the Industrial Doctorates Plan of the Secretariat of Universities and Research of the Department of Economy and Knowledge of the Generalitat of Catalonia.

\end{acknowledgement}

\begin{suppinfo}

Table S1: Detailed results of the retrospective validation analysis. \href{https://figshare.com/articles/SkeleDock_Table_S1/12248702}{(csv)} 
Table S2: PDB codes used as template for SkeleDock in the D3R GC4 pose prediction challenge. MD simulation: Description of the MD protocol used to refine the poses in D3R GC4 pose prediction challenge. Table S3: Mean RMSD obtained by the different methods used to model macrocycles. Table S4: Mean  RMSD  (computed  for  macrocycle  atoms  only)  obtained  by  the  different methods used to model macrocycles. \href{https://figshare.com/articles/SkeleDock_Supplementary_Information/12248693}{(pdf)}
\end{suppinfo}

\end{multicols}
\end{small}

\newpage 

\newpage 
\bibliography{references}
\newpage 
\end{document}